\documentclass[12pt]{article}
\usepackage[cp1251]{inputenc}
\usepackage[russian]{babel}
\usepackage{graphicx}

\def\baselinestretch{1.2}
\begin{document}
\begin{center}
{\Large \bf The value of long-range interactions parameter for some alloys} \vspace{0.5cm}

{\bf S.V.Belim\\
Omsk State University, Omsk, Russia\\
belim@univer.omsk.su}
\end{center}

\vskip 2mm \begin{center} \parbox{142mm} {\small
The critical behavior of some alloys are analyzed within the framework of Heisenbergs model with long-range interaction.
On based experimental values of the critical exponent $\gamma$ we calculate the value of paerameter of long-range interaction.
}
\end{center} \bigskip

\begin{center}
\end{center}

\section{Introduction}

In series of paper \cite{b1, b2, b3, b4, b5} experimental critical exponents are differ from results of theoretic-fields approach
for 3D models of Heisenberg ($\gamma=1.386$, $\beta=0.364$), 3D XY-model ($\gamma=1.316$, $\beta=0.345$) and 3D Ising
model ($\gamma=1.241$, $\beta=0.325$) \cite{b6}. Authors of these papers are  indicate that next-nearest-neighbor interaction
must be take into account for explanation the difference of experimental results from theoretic results.
This interaction may be take into account by means of term of hamiltonian of the form $J(r)\sim r^{-D-\sigma}$, where $D$ is
dimension of system and  $\sigma$ is parameter of long-range intaraction \cite{b7}.

In \cite{b1} the critical magnetic behavior of $EuO$ is investigated. Critical exponents of this system are $\gamma=1.29\pm 0.01$,
$\beta=0.368\pm 0.005$. In this article swoun that next-nearest-neighbor interaction $J_2$ must be take into account.
The  next-nearest-neighbor interaction is equal $(0.5\pm 0.2)J_1$ ($J_1$ -- nearest interaction).

In paper\cite{b2} critical exponents of $La_{0.5}Sr_{0.5}CoO_{3}$ are measured ($\gamma=1.351\pm 0.009$, $\beta=0.321\pm 0.002$).
Earlier in \cite{b3} the critical behavior was investigated in alloys $La_{1-x}Sr_{x}CoO_{3}$ ($0.2\leq x\leq 0.3$). Critical
exponents of these alloys have values $0.43\leq\beta\leq 0.46$, $1.39\leq \gamma\leq 1.43$.

The differ of critical exponents from theoretic results for short-range systems was finding for ferromagnetic phase transition
in $La_{0.1}Ba_{0.9}VS_{3}$ \cite{b4}. Critical exponents of these alloys have values $\gamma=1.366$, $\beta=0.501$.
Similar results were found in \cite{b5} for alloys $Fe_{90-x}Mn_xZr_{10}$ ($0\leq x \leq 16$).

Critical exponents of three-dimensional system with long-range intaraction for varios value $\sigma$ were calculate in \cite{b8}.
In this paper was shown that value of exponent $\gamma$ increase and exponent $\beta$ decrease with increasing
of parameter $\sigma$. If $\sigma$ is greater then $2$, then Heisenberg exponents are valid. If $\sigma$ is greater then $1.5$,
then mean feild exponents are valid. In interval $1.5<\sigma<2$ there are new classes of universality.

The aim of this paper is calculation value of parameter of long-range intaraction from experimental values of critical exponenets for varios systems.

\section{The theoretic-field description}

The Hamiltonian of a system with long-range effects can be written as
\begin{equation}\label{gam1}
   H=\int d^Dq\Big\{\frac{1}{2}(\tau_0+q^\sigma)\varphi^2+u_0\varphi^4\Big\},
\end{equation}
where $\varphi$ is the $n$-dimension order-parameter fluctuations, $D$ is the space dimensionality, $\tau_0\sim|T-T_c|$,
$T_c$ is the critical temperature, and $u_0$ is a positive constant. The critical behavior depends essentially on the parameter
$\sigma$ that determines the rate of interaction decay with increasing distance. As was shown in \cite{b7}, the influence of long-range
effects is appreciable for $0<\sigma<2$, while the critical behavior at $\sigma \geq 2$ is equivalent to the behavior of
short-range systems. For this reason, we restrict ourselves in what follows to the case $0<\sigma<2$.

The standard renormalization-group procedure based on the Feynman diagrams \cite{b6} with the
$G(\vec{k})=1/(\tau+|\vec{k}|^\sigma)$ propagator yields the following expressions for the functions
$\beta$,$\gamma_\varphi$ and $\gamma_t$, specifying the differential renormalization-group equation:
\begin{eqnarray}\label{beta}
    &&\beta=-(2\sigma-D)\Big[1-4(n+8)v+\Big(64(5n+22)(2\widetilde{J_1}-1)-128(n+2)\widetilde{G}\Big)v^2)\Big],\nonumber\\
    &&\gamma_t=(2\sigma-D)\Big(-2(n+2)v+48(n+2)\Big(2\widetilde{J_1}-1-\frac13\widetilde{G}\Big)v^2\Big), \\
    &&\gamma_\varphi=64(n+2)\widetilde{G}v^2,\nonumber\\
    &&v=u\cdot J_0,\ \ \ \widetilde{J_1}=\frac{J_1}{J_0^2}\ \ \ \  \widetilde{G}=\frac{G}{J_0^2}.\nonumber\\
    &&J_1=\int \frac{d^Dqd^Dp}{(1+|\vec{q}|^a)^2(1+|\vec{p}|^a)(1+|q^2+p^2+2\vec{p}\vec{q}|^{a/2})},\nonumber\\
    &&J_0=\int \frac{d^Dq}{(1+|\vec{q}|^a)^2},\nonumber\\
    &&G=-\frac{\partial}{\partial |\vec{k}|^a}\int \frac{d^Dq
    d^Dp}{(1+|q^2+k^2+2\vec{k}\vec{q}|^a)(1+|\vec{p}|^a)(1+|q^2+p^2+2\vec{p}\vec{q}|^{a/2})}\nonumber
\end{eqnarray}

The expressions obtained for the $\beta$-functions are asymptotic series and summation methods must be used to extract necessary
physical information from these series. In this work, the following Borel–Leroy transformation, which provides adequate results for series
appearing in the theory of critical phenomena \cite{b9}, is used
\begin{equation}
\begin{array}{rl}
  & f(v)=\sum\limits_{i}c_{i}v^{i}=\int\limits_{0}^{\infty}e^{-t}t^bF(vt)dt,  \\
  & F(v)=\sum\limits_{i}\frac{\displaystyle c_{i}}{\displaystyle(i+b)!}v^{i}.
\end{array}
\end{equation}
The [2/1] approximants with variation of the parameter $b$ are used to calculate the $\beta$-functions in
the two-loop approximation. As shown in \cite{b9}, such variation of $b$ makes it possible to determine the range
of variation of the vertex functions and to estimate the accuracy of the critical exponents obtained.

The critical behavior regime is fully determined by the stable fixed points of the renormalization-group transformation;
these points can be found from the condition that the $\beta$ functions vanish:
\begin{equation}\label{nep}
    \beta(v^*)=0.
\end{equation}
The condition for stability reduces to the requirement that the $\beta$-function derivative at the fixed point be positive:
\begin{equation}\label{proiz}
   \lambda=\frac{\partial\beta(v^*)}{\partial v}>0.
\end{equation}

The index $\nu$ characterizing the growth of correlation radius in the vicinity of critical point $(R_c\sim|T-T_c|^{-\nu})$ is
found from the expression:
\begin{eqnarray}
  \nu=v(\sigma+\gamma_t)^{-1}.\nonumber
\end{eqnarray}

The Fisher index $\eta$ describing the behavior of correlation function in the vicinity of critical point in the wave-vector
space $(G\sim k^{\sigma+\eta})$ is determined by the scaling function $\gamma_\varphi$: $ \eta=2-\sigma+\gamma_\varphi$.
Other critical indices can be determined from the scaling relations:
\begin{equation}
  \gamma=\nu(\sigma-\eta),\ \ \ \ \ \beta=\frac{\nu}{2}(D-\sigma+\eta).
\end{equation}

It is worth noting that the Pade–Leroy summation procedure is possible not for any $b$ values
and this significantly limits the possibility of applying the method. This limitation is associated with the
appearance of the poles of the approximants near the solutions of the system of Eqs. (\ref{beta}); for this reason, it
is impossible to determine the position of the fixed points. In this work, the parameter $b$ varies from $0$ to a
value beginning with which the determination of the stable fixed point becomes impossible. In this range, $20$ values of the parameter
$b$ are taken for which the fixed points are searched. Average values with a certain accuracy determined by the spread in the values for various
$b$ values are taken as the effective charges at the fixed point.

\section{The value of parameter $\sigma$}

Lets we consider the applicability models with long-range interaction for explanation of experimental data.
The value of parameter $\sigma$ shell calculeted on based of experimental value of critical exponent $\gamma$.

For $EuO$ \cite{b1} exprimental critical exponents have values $\gamma=1.29\pm 0.01$, $\beta=0.368\pm 0.005$. Within the framework of
Heisenbergs model ($n=3$) the value of critical exponent $\gamma=1.290\pm 0.002$ correspond the value $\sigma=1.941$ and
critical exponent $\beta=0.376\pm 0.008$. Within the framework of XY-model ($n=2$) the value of critical exponent $\gamma=1.29\pm 0.03$
correspond the value $\sigma=1.991$ and critical exponent $\beta=0.354\pm 0.007$. As can be seen from comparison theoretical and experimental
results the Heisenbergs model with long-range interaction demonstrate satisfactory conformance to experiment.

For $La_{0.5}Sr_{0.5}CoO_{3}$ \cite{b2} exprimental critical exponents have values $\gamma=1.351\pm 0.009$, $\beta=0.321\pm 0.002$.
Within the framework of Heisenbergs model ($n=3$) the value of critical exponent $\gamma=1.351\pm 0.002$ correspond the value $\sigma=1.980$ and
critical exponent $\beta=0.368\pm 0.004$. Within the framework of XY-model ($n=2$) the value of critical exponent $\gamma=1.351$ don't exist,
the maximum value of $\gamma$ is equal $1.316$ for $\sigma=2$. As can be seen the differ theoretical results from experimental results is significant.
However in \cite{b2} the critical exponent $\gamma$ was measured for the critical temperature $T_c=223.18\ K$, and the critical
exponent $\beta$ for the critical temperature $T_c=222.82\ K$. As is well known that values of critical exponents are very strong depends
from selection of the critical temperature.

For $La_{0.1}Ba_{0.9}VS_{3}$ \cite{b4} exprimental critical exponents have values $\gamma=1.366$, $\beta=0.501$.
Within the framework of Heisenbergs model ($n=3$) the value of critical exponent $\gamma=1.366\pm 0.002$ correspond the value $\sigma=1.990$
and critical exponent $\beta=0.369\pm 0.009$. In this case the XY-model are not valid. It is shown that theoretical value of critical
exponent $\beta$ don't agree the result of experiment. But the experimental value $\beta=0.501$ is very strange, because limit $\beta\leq 0.5$
is always valid. The equality $\beta=0.5$ is valid only for mean field theory, in which $\gamma=1$.

Critical exponents of alloys $Fe_{90-x}Mn_xZr_{10}$ ($0\leq x \leq 16$ depend on parameter $x$ \cite{b5}. In the table there are experimental
values of critical exponents from paper \cite{b5}. Also in this table there are parameters $\sigma$ and critical exponents $\beta_H$,
which calculated Within the framework of Heisenbergs model with long-range interaction on the grounds of values of critical exponents $\gamma$.
\begin{center}
\begin{tabular}{|c|c|c|c|c|}
  \hline
  $x$& $\gamma$ & $\beta$ & $\sigma$ & $\beta_H$ \\
  \hline
  0 & 1.376 & 0.369 & 1.995 & $0.36\pm 0.01$ \\
  4 & 1.383 & 0.373 & 1.998 & $0.37\pm 0.01$ \\
  6 & 1.359 & 0.358 & 1.984 & $0.37\pm 0.02$ \\
  8 & 1.364 & 0.355 & 1.987 & $0.36\pm 0.02$ \\
  10& 1.406 & 0.356 &  -    &  -\\
  12& 1.395 & 0.376 &  -    &  -\\
  16& 1.412 & 0.362 &  -    &  -\\ \hline
\end{tabular}
\end{center}
For values $x>8$ the parameter $\sigma$ don't calculated, because the maximum value of exponent $\gamma$ is $1.386$ for value parameter $\sigma=2$.
It is shown that in interval $0\leq x\leq 8$ the theory demonstrate agreement with experiment.

As can be seen from calculation the use of Heisenbergs model with long-range interaction is valid for explanation of experimental data.

\section{Acknowledgments}

This work was supported by the Russian Foundation for Basic Research (project no. 06-02-16018).

\def\baselinestretch{1.0}

\end{document}